# Even-Balanced States Excitation in Two-Qubits System

Yoav Koral[1], Shilo Avraham[3], Manimuthu Periyasamy[1], Guy Deutscher[3], Shmuel E. Schacham[1], and Eliyahu Farber[2]

[1]Department of Electrical and Electronic Engineering, Ariel University, P.O.B. 3, Ariel 40700

[2]Department of Electrical and Electronic Engineering and Department of Physics, Ariel University, P.O.B. 3, Ariel 40700

[3] Raymond and Beverly Sackler School of Physics and Astronomy, Tel Aviv University, Tel Aviv, Israel

*Abstract* – **An essential step in building a quantum computer system is combining two qubits. Previous theoretical analyses of this element have been based primarily on the short schematic matrix approaches. In the present work we extend the investigation to the combination of two qubits by implementing the Dirac picture method as a theoretical framework. We conduct simulations, using parameters of grAl qubits, to demonstrate the feasibility of driving two qubits into an even-balanced quantum state. The research highlights the effectiveness of the Dirac picture method in achieving the desired state, enabling improved control of entangled states and advanced quantum computing research. The simulation results and the theoretical calculations agree within of 0.2%, and the two examples at section IV, show a good agreement to the level of 6% error from the even balanced state.**

## I. INTRODUCTION

The quantum computer is widely regarded as the most effective tool for parallel computing. It shows a great potential for efficient parallel 3D operations using a minimal number of qubits, surpassing the capabilities of conventional Graphic Processing Units [1]. Advances in the investigation and implementation of quantum computers have greatly improved the understanding of quantum states and their utilization. An essential requirement for efficient quantum computing operation is obtaining an initial even-balanced state in the Two-Qubit circuit.

Even though the incorporation of qubits into a processor received extensive theoretical investigation, a thorough research of the two-qubit circuit is essential for the advancement of quantum computer. Achieving a precise control over the quantum states of two qubits is a crucial step for fully harnessing the potential of quantum computers. In previous research, like the work done by Kranz et al. [2] , important findings were obtained by examining the characteristics of single qubits and how they can be used in practice. However, moving from individual qubits to coupled qubit systems creates new challenges that require creative solutions. There are different theoretical frameworks for explaining the dynamics of qubit coupling, from straightforward schematic matrix methods to advanced analysis based on the Dirac picture. Although schematic matrix methods are simple and computationally efficient, they may not fully capture the complexities of qubit interactions. Some researchers discuss only the initiation of the Bell states [3, 4]. Others deal with a general description of gate operations [5-7] as well as special cases of atoms coupling [8] and the entanglement function [9], or with a weak coupling case [10]. Despite significant progress, achieving accurate control over two-qubit systems remains a challenging task.

The Dirac picture method is a reliable tool for understanding the quantum dynamics of coupled qubit systems. Using this method, our research aims to explore the creation of even-balanced states in two-qubit systems. By combining theoretical analysis with computational simulations, we intend to enhance the understanding of the mechanisms driving the transition to the desired state. We utilize a Two-Qubit Hamiltonian simulation, using the Dirac picture method, to thoroughly analyze the excitation response and describe in detail the process of obtaining an even-balanced state from the initial |00⟩ ground state.

Following the theoretical analysis, we present the simulation process of two qubit excitations in Chapter II. The end state approximation is described in Chapter III. Following this analysis, we show in Chapter IV examples of simulation results, based on our experimental results obtained from Granular aluminum (grAl) films, which is our preferred choice for implementing qubits, due to their characteristics, such as high kinetic induction and minimal losses near to the Mott transition [11-16].

## II. SIMULATION PROCESS

A Two-Qubits circuit with capacitive coupling is shown in Fig. 1.

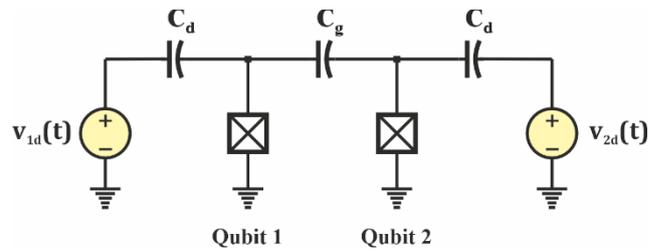

Fig. 1. Two-Qubits system with capacitive coupling

This circuit introduces two qubits that are coupled by a gate capacitor and driven by two voltage sources. Each qubit can be depicted as a parallel LC network, ignoring its non-linear properties, as shown in Fig. 2.

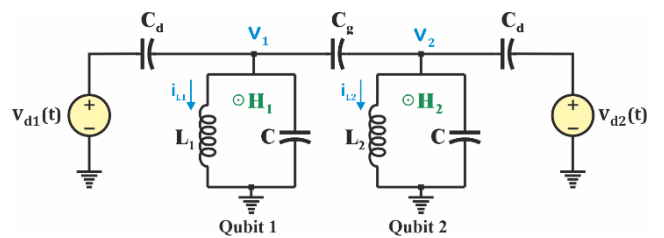

Fig. 2: Two-Qubits circuit, Linear Model.

The theoretical analysis of this circuit begins with the Kirchhoff current equations set:

$$C_d \frac{d}{dt}(V_{d1} - v_1) - C_g \frac{d}{dt}(v_1 - v_2) = C_J \frac{dv_1}{dt} + i_{L_1}$$
$$C_d \frac{d}{dt}(V_{d2} - v_2) + C_g \frac{d}{dt}(v_1 - v_2) = C_J \frac{dv_2}{dt} + i_{L_2}$$
(1)

where

$$i_{L_i} = \phi_i/L_i, \qquad \dot{\phi}_i = v_i \qquad i = 1,2 \quad (2)$$

here $\phi_i, L_i, i_{L_i}$ and $v_i$ are the inductors magnetic flux, inductance, current, and voltage, respectively. Using the Euler-Lagrange formulation and the Legendre transformation [17], together with the flux to junction-charge relation with short voltage sources:

$$\begin{bmatrix} q_1 \\ q_2 \end{bmatrix} = \begin{bmatrix} C_s & -C_g \\ -C_g & C_s \end{bmatrix} \begin{bmatrix} \dot{\phi}_1 \\ \dot{\phi}_2 \end{bmatrix} \; ; \; C_s = C_J + C_d + C_g \quad (3)$$

we obtain the Hamiltonian of the circuit in a quantum operator form, as follows:

$$H = \frac{\Phi_1^2}{2L_1} + \frac{Q_1^2}{2C_M} + \frac{\Phi_2^2}{2L_2} + \frac{Q_2^2}{2C_M} + \frac{Q_1 Q_2}{C_G}$$
$$+ (V_{d1}\mu_d + V_{d2}\mu_g)Q_1 + (V_{d1}\mu_g + V_{d2}\mu_d)Q_2$$
(4)

where

$$\gamma = \frac{C_g}{C_s}$$
$$C_M = C_s(1 - \gamma^2) \quad ; \quad C_G = C_s\left(\frac{1}{\gamma} - \gamma\right)$$
$$\mu_d = \frac{C_d}{C_s} \cdot \frac{1}{1 - \gamma^2} \quad ; \quad \mu_g = \frac{C_d}{C_s} \cdot \frac{\gamma}{1 - \gamma^2}$$
(5)

Normalizing this Hamiltonian [18], we get:

$$H = \omega_1\left(\tilde{a}_1^\dagger \tilde{a}_1 + \frac{1}{2}\right) + \omega_2\left(\tilde{a}_2^\dagger \tilde{a}_2 + \frac{1}{2}\right)$$
$$- \omega_g(\tilde{a}_1^\dagger - \tilde{a}_1)(\tilde{a}_2^\dagger - \tilde{a}_2)$$
$$+ i(\Omega_{d1d} + \Omega_{d2g})(\tilde{a}_1^\dagger - \tilde{a}_1)$$
$$+ i(\Omega_{d1g} + \Omega_{d2d})(\tilde{a}_2^\dagger - \tilde{a}_2)$$
(6)

where

$$i = 1,2 \quad ; \quad \omega_i = \frac{1}{\sqrt{L_i C_M}}$$
$$\Omega_{d1d} = \frac{\mu_d}{\sqrt{2}\sigma_{\phi 1}} V_{d1} \quad ; \quad \Omega_{d2g} = \frac{\mu_g}{\sqrt{2}\sigma_{\phi 1}} V_{d2}$$
$$\Omega_{d1g} = \frac{\mu_g}{\sqrt{2}\sigma_{\phi 2}} V_{d1} \quad ; \quad \Omega_{d2d} = \frac{\mu_d}{\sqrt{2}\sigma_{\phi 2}} V_{d2}$$
$$\omega_g = \frac{\gamma}{2}\sqrt{\omega_1 \omega_2} \quad ; \quad \sigma_{\phi i} = \sqrt{\hbar\sqrt{L_i/C_M}}$$
(7)

and the extended annihilation and creation operators [19] are

$$\tilde{a}_1 = a \otimes \mathbb{I} \quad ; \quad \tilde{a}_2 = \mathbb{I} \otimes a$$
$$\tilde{a}_1^\dagger = a^\dagger \otimes \mathbb{I} \quad ; \quad \tilde{a}_2^\dagger = \mathbb{I} \otimes a^\dagger$$
(8)

where $a$ is the annihilation operator of linear harmonic oscillator (B.1.b in [18]), and its adjoint operator $a^\dagger$ is the creation operator. $\mathbb{I}$ is the identity matrix with the same size as $a$, $\otimes$ is the Kronecker product operator. Assuming the qubit is of Fluxonium type, where the energy level of $|2\rangle$ state is higher than that of the states $|0\rangle$ and $|1\rangle$, it is reasonable to ignore the higher energy levels in our calculations and reduce the sizes of $a$ and $\mathbb{I}$ matrices to 2x2. The Hamiltonian then gets the form:

$$H \cong H_0 + H_d$$
$$H_0 = -\frac{\omega_1}{2}\tilde{\sigma}_{z1} - \frac{\omega_2}{2}\tilde{\sigma}_{z2} + \omega_g \tilde{\sigma}_{y1}\tilde{\sigma}_{y2}$$
$$H_d = (\Omega_{d1d} + \Omega_{d2g})\tilde{\sigma}_{y1} + (\Omega_{d1g} + \Omega_{d2d})\tilde{\sigma}_{y2}$$
(9)

where $\tilde{\sigma}_{yi}$ and $\tilde{\sigma}_{zi}$ are the extended Pauli matrices [20]. The driving signals are assumed to be sufficiently small so that none of the higher states, starting from $|2\rangle$ and above, are being excited. Solving the entire Hamiltonian directly will produce a solution that is a mix of the eigen oscillation and the excitation response. This will compel us to make an additional separation process. Therefore, to tackle the excitation response only, we will use the Dirac picture simulation to solve the Hamiltonian in Eq. (**9**). Using this method, we can use the simpler calculation of the eigen-functions of the Hamiltonian's static part $H_0$, to determine the solution for the entire Hamiltonian. This is done by calculating the rotating frame unitary matrix, using it to derive the rotating frame transformation of the Drive Hamiltonian. Based on this result, using a standard algebraic process, we calculate the required evolution matrix. This procedure is summarized by the following process [2]:

$$U_{rf} = e^{iH_0 t} \qquad (a)$$
$$\tilde{H}_d = U_{rf} H_d U_{rf}^\dagger \qquad (b)$$
$$\tilde{H}_d \xrightarrow{Diagonalization} \{\Lambda_d, P_d\} \qquad (c)$$
$$\Theta_d = \int_0^t \Lambda_d(t')\, dt' \qquad (d)$$
$$U_d = P_d\, e^{-i\Theta_d(t)} P_d^\dagger \qquad (e)$$
$$\psi_f = U_d \psi_0 \qquad (f)$$
(10)

where $U_{rf}$ is the Rotating Frame unitary matrix, $\tilde{H}_d$ is the Rotated Frame Drive Hamiltonian created by the Dirac picture formula, $U_d$ is the Rotated Frame Drive Evolution matrix, and $\psi_0$ and $\psi_f$ are the initial and the end states, respectively.

## III. END STATE APPROXIMATION

### A. Identical Qubits

In order to check the validity of the simulation, it is necessary to do comparative analysis. To simplify the calculation, we assume that the resonators are equal, $L = L_1 = L_2$. The Hamiltonian for this case becomes

$$H \cong H_0 + H_d$$
$$H_0 = -\frac{\omega_r}{2}\widetilde{\sigma}_{z1} - \frac{\omega_r}{2}\widetilde{\sigma}_{z2} + \omega_{gr}\widetilde{\sigma}_{y1}\widetilde{\sigma}_{y1} \quad (11)$$
$$H_d = (\Omega_{d1d} + \Omega_{d2g})\widetilde{\sigma}_{y1} + (\Omega_{d1g} + \Omega_{d2d})\widetilde{\sigma}_{y2}$$

where

$$\omega_r = \frac{1}{\sqrt{LC_M}} \quad ; \quad \omega_{gr} = \frac{\gamma}{2}\omega_r \quad (12)$$

### B. Even Excitation

We start the excitation analysis with a simplified circuit in which the voltage sources are equal, i.e. $V_d = V_{d1} = V_{d2}$. For this case, the Hamiltonian in Eq. (11) can be simplified as follows:

$$H_e \cong H_0 + H_d$$
$$H_0 = -\frac{\omega_r}{2}\widetilde{\sigma}_{z1} - \frac{\omega_r}{2}\widetilde{\sigma}_{z2} + \omega_{gr}\widetilde{\sigma}_{y1}\widetilde{\sigma}_{y2} \quad (13)$$
$$H_d = \Omega_{de}(\widetilde{\sigma}_{y1} + \widetilde{\sigma}_{y2})$$

where

$$\Omega_{de} = \frac{\mu_d + \mu_g}{\sqrt{2}\,\sigma_\phi}V_d \quad (14)$$

The Hamiltonian in Eq. (13) is composed of two parts. The first part, $H_0$ represents the Circuit Hamiltonian which describes the electrical circuit.

$$H_0 = \begin{bmatrix} -\omega_r & 0 & 0 & -\omega_{gr} \\ 0 & 0 & \omega_{gr} & 0 \\ 0 & \omega_{gr} & 0 & 0 \\ -\omega_{gr} & 0 & 0 & \omega_r \end{bmatrix} \quad (15)$$

The second part, $H_d$ is the Drive Hamiltonian, which describes the driving signal.

$$H_d = i\,\Omega_d \begin{bmatrix} 0 & -1 & -1 & 0 \\ 1 & 0 & 0 & -1 \\ 1 & 0 & 0 & -1 \\ 0 & 1 & 1 & 0 \end{bmatrix} \quad (16)$$

Next, using Eq. (15), we get the Rotating Frame unitary matrix:

$$U_{rf} = e^{i\widetilde{H}_0 t} = \begin{bmatrix} f_{\Sigma+} & 0 & 0 & -ig_\Sigma \\ 0 & f_\Delta & ig_\Delta & 0 \\ 0 & ig_\Delta & f_\Delta & 0 \\ -ig_\Sigma & 0 & 0 & f_{\Sigma-} \end{bmatrix} \quad (17)$$

where:

$$\lambda_\Delta = \frac{\gamma}{2}\omega_r \quad ; \quad \lambda_\Sigma = \sqrt{1+\frac{\gamma^2}{4}}\cdot\omega_r \quad (18)$$
$$\beta_\Sigma = \pi - \arctan\frac{\gamma}{2}$$
$$f_{\Sigma\pm} = \cos\lambda_\Sigma t \pm i\sin\lambda_\Sigma t\,\cos\beta_\Sigma \quad ; \quad g_\Sigma = \sin\lambda_\Sigma t\,\sin\beta_\Sigma$$
$$f_\Delta = \cos\lambda_\Delta t \quad ; \quad g_\Delta = \sin\lambda_\Delta t$$

By substituting Eq. (15) and Eq. (17) into Eq. (10a) of the Dirac picture formula, we obtain the Rotating Drive Hamiltonian:

$$\widetilde{H}_d = \begin{bmatrix} 0 & h_a^* & h_a^* & 0 \\ h_a & 0 & 0 & h_b^* \\ h_a & 0 & 0 & h_b^* \\ 0 & h_b & h_b & 0 \end{bmatrix}$$

$$h_a = \Omega_{de}\,e^{i\lambda_\Delta t}\left[\sqrt{2}\cos\left(\frac{\pi}{4}-\beta_\Sigma\right)\sin\lambda_\Sigma t + i\cos\lambda_\Sigma t\right] \quad (19)$$

$$h_b = \Omega_{de}\,e^{-i\lambda_\Delta t}\left[\sqrt{2}\sin\left(\frac{\pi}{4}-\beta_\Sigma\right)\sin\lambda_\Sigma t + i\cos\lambda_\Sigma t\right]$$

Next, we set the relative voltage drive of Eq. (19) in the form of a sinusoidal waveform with a Gaussian shape, which is centered at the $\lambda_\Sigma$ eigen frequency.

$$\Omega_{de} = \Omega_e\,e^{-\frac{1}{2}\left(\frac{t}{\sigma_t}\right)^2}\cdot\sin(\lambda_\Sigma t - \varphi_d) \quad (20)$$

where $\Omega_e$ is the Gaussian peak of the relative voltage, $\sigma_t$ is the Gaussian standard deviation, and $\varphi_d$ is the sinusoidal phase shift due to the Gaussian peak. By using the Rotating Wave Approximation (RWA) we remove the high frequency product. By multiplying Eq. (20) with Eq. (19), we obtain the following expressions

$$h_a \triangleq -\Omega_e\,\rho_C\,e^{-\frac{1}{2}\left(\frac{t}{\sigma_t}\right)^2}e^{-i\lambda_\Delta t}$$
$$h_b \triangleq -\Omega_e\,\rho_S\,e^{-\frac{1}{2}\left(\frac{t}{\sigma_t}\right)^2}e^{i\lambda_\Delta t} \quad (21)$$

where

$$\rho_C = \frac{1}{\sqrt{2}}\cos\left(\frac{\pi}{4}-\beta_\Sigma\right)$$
$$\rho_S = \frac{1}{\sqrt{2}}\sin\left(\frac{\pi}{4}-\beta_\Sigma\right) \quad (22)$$

The $\triangleq$ sign denotes the fact that these expressions include only the terms which are dependent on the frequency difference. We omitted the terms dependent on the sum of the frequencies since their contributions to the integral (10d) are negligible as compared with those of the frequency difference.

Using Eq. (5) and Eq. (14), we get:

$$\Omega_e = \frac{C_d}{C_s(1-\gamma)}\cdot\frac{V_s}{\sigma_\phi} \quad (23)$$

Next, we substitute Eq. (19) in Eq. (10), and we obtain the evolution matrix for the even excitation case:

$$U_{de} = \begin{bmatrix} 1 - 4\rho_C^2 \sin^2\frac{\theta_e}{4} & i\rho_C \sin\frac{\theta_e}{2} & i\rho_C \sin\frac{\theta_e}{2} & -4\rho_C\rho_S \sin^2\frac{\theta_e}{4} \\ i\rho_C \sin\frac{\theta_e}{2} & \cos^2\frac{\theta_e}{4} & -\sin^2\frac{\theta_e}{4} & i\rho_S \sin\frac{\theta_e}{2} \\ i\rho_C \sin\frac{\theta_e}{2} & -\sin^2\frac{\theta_e}{4} & \cos^2\frac{\theta_e}{4} & i\rho_S \sin\frac{\theta_e}{2} \\ -4\rho_C\rho_S \sin^2\frac{\theta_e}{4} & i\rho_S \sin\frac{\theta_e}{2} & i\rho_S \sin\frac{\theta_e}{2} & 1 - 4\rho_S^2 \sin^2\frac{\theta_e}{4} \end{bmatrix} \quad (24)$$

and the shift angle is:

$$\frac{\theta_e}{2} = \sqrt{2\pi} \, \frac{C_d}{C_s(1-\gamma)} \, \sigma_t e^{-\frac{1}{2}\lambda_\Delta^2 \sigma_t^2} \cdot \frac{V_s}{\sigma_\phi} \quad (25)$$

Consequently, the end state for the even excitation state from the $|00\rangle$ state is:

$$\psi_0 = \begin{bmatrix} 1 \\ 0 \\ 0 \\ 0 \end{bmatrix} \Rightarrow \psi_e = \begin{bmatrix} 1 - 4\rho_C^2 \sin^2\frac{\theta_e}{4} \\ i\rho_C \sin\frac{\theta_e}{2} \\ i\rho_C \sin\frac{\theta_e}{2} \\ -4\rho_C\rho_S \sin^2\frac{\theta_e}{4} \end{bmatrix} \quad (26)$$

To obtain the approximation of the end state, the limits of the integral for the shift angle $\theta_e$ are approximated to be between the $-\infty$ and $+\infty$ [21]. The same approximation is applied to $\theta_o$ below.

### C. Odd Excitation

In a similar manner to Eq. (13), we assume that for the odd excitation case, $V_d = V_{d1} = -V_{d2}$, and its Hamiltonian from Eq. (11) is simplified to:

$$H_o \cong H_0 + H_d$$
$$H_0 = -\frac{\omega_r}{2}\tilde{\sigma}_{z1} - \frac{\omega_r}{2}\tilde{\sigma}_{z2} + \omega_{gr}\tilde{\sigma}_{y1}\tilde{\sigma}_{y2} \quad (27)$$
$$H_d = \Omega_{do}(\tilde{\sigma}_{y1} - \tilde{\sigma}_{y2})$$

where:

$$\Omega_{do} = \frac{\mu_d - \mu_g}{\sqrt{2}\,\sigma_\phi} V_d \quad (28)$$

Further, in a similar manner to the process in section III.B, we obtain the evolution matrix for the odd excitation case:

(29)

$$U_{do} = \begin{bmatrix} 1 - 4\rho_S^2 \sin^2\frac{\theta_o}{4} & i\rho_S \sin\frac{\theta_o}{2} & -i\rho_S \sin\frac{\theta_o}{2} & 4\rho_C\rho_S \sin^2\frac{\theta_o}{4} \\ i\rho_S \sin\frac{\theta_o}{2} & \cos^2\frac{\theta_o}{4} & \sin^2\frac{\theta_o}{4} & -i\rho_S \sin\frac{\theta_o}{2} \\ -i\rho_S \sin\frac{\theta_o}{2} & \sin^2\frac{\theta_o}{4} & \cos^2\frac{\theta_o}{4} & i\rho_S \sin\frac{\theta_o}{2} \\ 4\rho_C\rho_S \sin^2\frac{\theta_o}{4} & -i\rho_C \sin\frac{\theta_o}{2} & i\rho_C \sin\frac{\theta_o}{2} & 1 - 4\rho_C^2 \sin^2\frac{\theta_o}{4} \end{bmatrix}$$

where the shift angle is

$$\frac{\theta_o}{2} = \sqrt{2\pi} \, \frac{C_d}{C_s(1+\gamma)} \, \sigma_t e^{-\frac{1}{2}\lambda_\Delta^2 \sigma_t^2} \cdot \frac{V_s}{\sigma_\phi} \quad (30)$$

Consequently, the end state for the odd excitation from the $|00\rangle$ state is:

$$\psi_0 = \begin{bmatrix} 1 \\ 0 \\ 0 \\ 0 \end{bmatrix} \Rightarrow \psi_o = \begin{bmatrix} 1 - 4\rho_S^2 \sin^2\frac{\theta_o}{4} \\ i\rho_S \sin\frac{\theta_o}{2} \\ -i\rho_S \sin\frac{\theta_o}{2} \\ 4\rho_C\rho_S \sin^2\frac{\theta_o}{4} \end{bmatrix} \quad (31)$$

## IV. SIMULATIONS EXAMPLES

### A. Initial State

We consider the ground state of the circuit, $|00\rangle$, as the initial state for the sake of convenience. Its wave function distribution is shown in Fig. 3. The simulations examine the state transfer from the initial $|00\rangle$ state to the end state, using the Dirac picture method. We examine both even and odd excitation in order to obtain a balanced state. Two examples of this process, presenting the excitation response only, are discussed. In all simulations we use the parameters of granular aluminum, based on our extensive experience with implementing qubits in grAl.

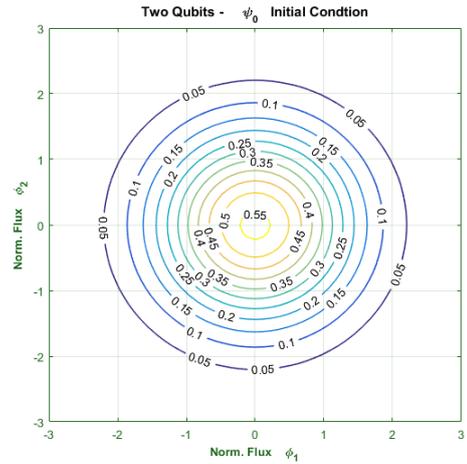

Fig. 3. Initial state $|00\rangle$ – Wave Function Distribution

### B. Even Excitation Case

The two signals for even excitation intended to achieve the transfer from the initial state to the even balanced state, $\psi_f = 0.5 \cdot [1,1,1,1]$ are shown in Fig. 4. We use the parameters of granular aluminum.

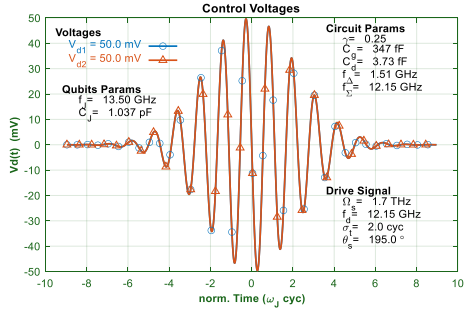

Fig. 4. Even excitation signals

The magnitude of the resultant coefficient $c_{ij}$ of the state wavefunction:

$$\psi(t) = c_{00}|00\rangle + c_{01}|01\rangle + c_{10}|10\rangle + c_{11}|11\rangle \quad (32)$$

undergoes a change during the excitation process as seen in Fig. 5:

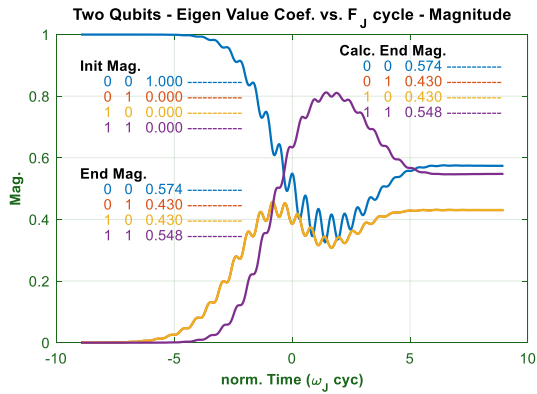

Fig. 5. Even excitation state response during the process.

For illustration purposes, the end state wave function distribution is shown at Fig. 6

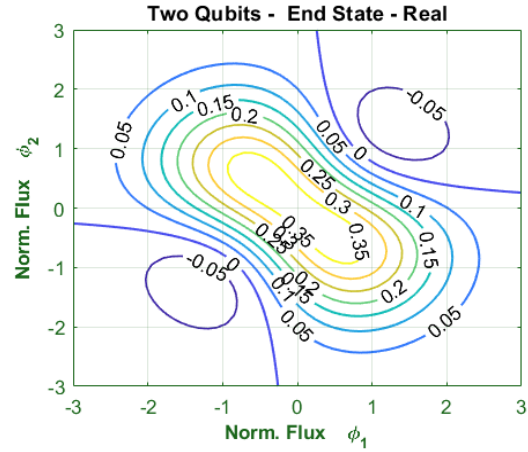

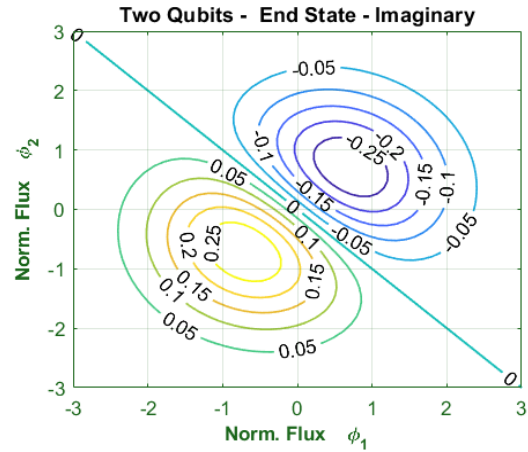

Fig. 6 Even excitation end state, $\psi_f \cong 0.5 \cdot [1,1,1,1]$

A good agreement between the simulated and the calculated end state magnitude is as shown in Fig. 5. The resulting RMS error for the end state magnitude, is calculated to be 0.066.

*C. Odd Excitation Case*

The two signals for odd excitation, intended to achieve the transfer from the initial state to the even balanced state, $\psi_f = 0.5 \cdot [1,1,1,1]$ are shown in Fig. 6.

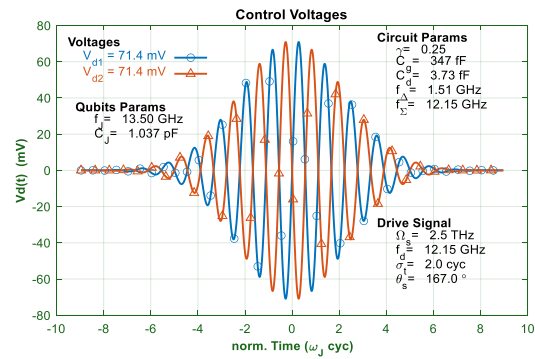

Fig. 7. Odd excitation signals

The magnitude of the resultant coefficient undergoes a change during the process as shown in Fig. 7.

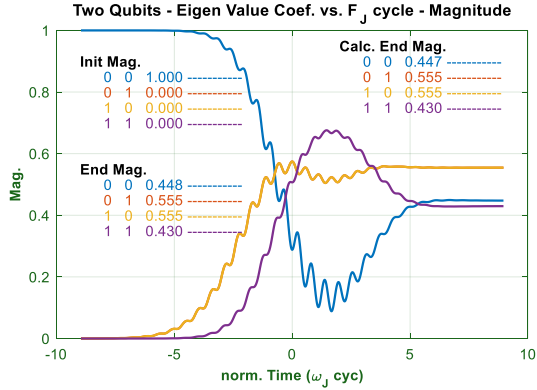

Fig. 8.  Odd excitation state response during excitation process

The end state wave function distribution, is at Fig. 8

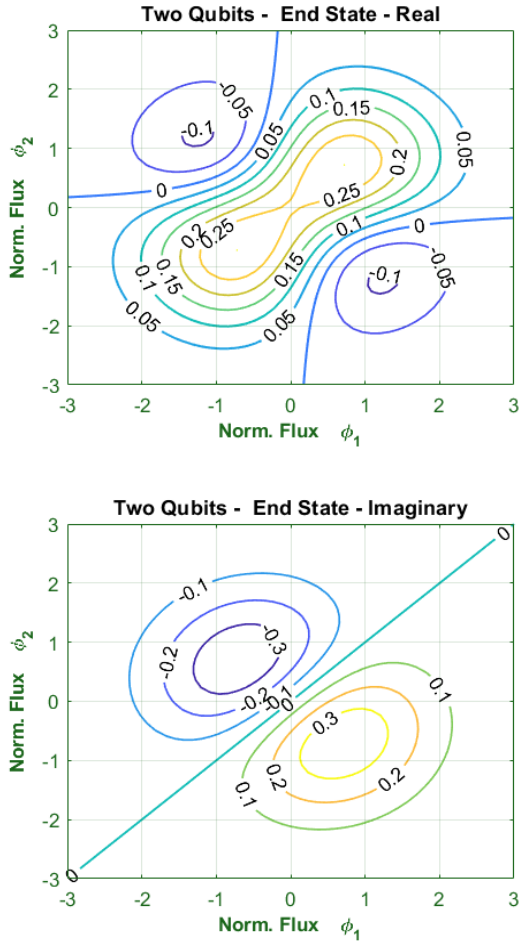

Fig. 9  Odd excitation end state : $\psi_f \cong 0.5 \cdot [1,1,1,1]$

A good agreement between the simulated and the calculated final magnitude can be observed in Fig. 8. The resulting RMS error for the end state magnitude, is calculated to be 0.059.

## V.  CONCLUSIONS

We introduced simulations of Two-Qubits circuit using the Dirac picture method. The simulation results agree well with the theoretical calculations, showing an error which does not exceed 0.2%. Furthermore, the two examples referring to the even and odd excitations, show a good agreement as well, with an average error of about only 6% from the desired outcome of even balanced state. These results provide a significant contribution to the ongoing development of scalable and efficient quantum computing technologies.